\documentclass[11pt,journal,twocolumn]{IEEEtran}
\IEEEoverridecommandlockouts

\usepackage{amsmath,amssymb,amsfonts,epsf,epsfig,times}
\usepackage[all]{xy}
\usepackage{graphicx,color}
\usepackage{subfigure}
\usepackage{url}
\usepackage{cite}
\usepackage{tikz}
\usepackage{epstopdf}

\usepackage{keyval,times}

\newtheorem{theorem}{Theorem}[section]
\newtheorem{lemma}{Lemma}[section]
\def\proof{\noindent{\it Proof: }}
\def\QED{\mbox{\rule[0pt]{1.5ex}{1.5ex}}}
\def\endproof{\hspace*{\fill}~\QED\par\endtrivlist\unskip}
\newcommand{\re}{\mathbb{R}}

\newcommand{\norm}[1]{\left\|#1\right\|}

\newcommand{\abs}[1]{\left|#1\right|}

\newcommand{\defeq}{\stackrel{\triangle}{=}}

\newtheorem{definition}[theorem]{Definition}

\newtheorem{myremark}[theorem]{Remark}

\newcommand{\Vcal}{\mathcal{V}}

\newcommand{\OMIT}[1]{}

\begin{document}
\title{UAV Circumnavigating an Unknown Target Under a GPS-denied Environment with Range-only Measurements}
\author{Yongcan Cao
\thanks{The author is with the Control Science Center of Excellence, Aerospace Systems Directorate, Air Force Research Laboratory, Wright-Patterson AFB, OH 45433, USA (email: yongcan.cao@gmail.com). Approved for public release; distribution unlimited, 88ABW-2014-0095.}
}

\markboth{}
         {}

\maketitle

\begin{abstract}
One typical application of unmanned aerial vehicles is the intelligence, surveillance, and reconnaissance mission, where the objective is to improve situation awareness through information acquisition. For examples, an efficient way to gather information regarding a target is to deploy UAV in such a way that it orbits around this target at a desired distance. Such a UAV motion is called \textit{circumnavigation}. The objective of the paper is to design a UAV control algorithm such that this circumnavigation mission is achieved under a GPS-denied environment using range-only measurement. The control algorithm is constructed in two steps. The first step is to design a UAV control algorithm by assuming the availability of both range and range rate measurements, where the associated control input is always bounded. The second step is to further eliminate the use of range rate measurement by using an estimated range rate, obtained via a sliding-mode estimator using range measurement, to replace actual range rate measurement. Such a controller design technique is applicable in the control design of other UAV navigation and control missions under a GPS-denied environment.
\end{abstract}


\IEEEpeerreviewmaketitle

\section{Introduction}

As Unmanned Aerial Vehicles (UAVs) gain more and more favor in both military and civilian applications due to its advantages over manned aircraft, it is now a demanding technology that UAVs can be used to accomplish missions with minimum human supervision. In many cases, complete autonomy is often desired in UAV operations unless human supervision is necessary. Technologies are thus needed to increase autonomy for UAV operations~\cite{Dahm10}.


The need of autonomy for UAV operations comes from two perspectives: performance and cost. From the performance's point of view, UAV with autonomy capability is more reliable than manned aircraft. Human operators are one of the most common sources of errors in complex systems. In many situations, human operators are more likely to make mistakes. It is also well-known that efficiency of human operators is expected to decrease after a certain period of time. All these drawbacks can be addressed if autonomy becomes available to replace human operators. From the cost's point of view, UAV with autonomy is less expensive than manned aircraft as the cost of training and/or replacing a pilot is high. Another factor of cost is that manned aircraft are generally larger in size and more complex in capabilities than UAVs due to the existence of human operators onboard.

Due to FAA regulations, UAVs are now mainly used in military and security applications, such as border patrol~\cite{Bolkcom04}, mapping~\cite{NagaiCSKA09}, and surveillance~\cite{QuigleyGGEB05}. For instance, UAVs can be used to gather information of a target by orbiting around it at some desired distance. Such a UAV mission is often called \textit{circumnavigation}. If GPS is available, both target location and UAV location can be obtained and this circumnavigation mission can be accomplished by using existing orbiting algorithms~\cite{BeardMcLain12}. However, the vulnerability of UAVs to GPS jamming and spoofing poses a significant threat to the safety of UAV operation. Recent tests confirm that GPS can be denied due to jamming~\cite{Warwick11} or spoofing~\cite{ShepardBhattiHumphreys12}. In military operations, GPS is more often jammed due to contested environment under which UAVs are operated. Refs.~\cite{ShamesDFA12,DeghatSAY13,MatveevTeimooriSavkin09,MatveevTeimooriSavkin11,CaoMCK13} were devoted to achieve circumnavigation under a GPS-denied environment. A localization-and-control framework was first proposed in~\cite{ShamesDFA12,DeghatSAY13} to solve the circumnavigation problem. The main idea is to first estimate the location of the target based on the location of the UAV under some local coordinate frame and then design controllers based on estimated target location and additional bearing/range measurement. Under a GPS-denied environment, the location of the UAV can be tracked down using inertial sensors at the cost of integration drift. An increasing drift means performance degradation. It is thus desired that the location of the UAV is not used in the controller design. A different control strategy was developed in~\cite{MatveevTeimooriSavkin09,MatveevTeimooriSavkin11} where both range and range rate measurements are used to design control algorithms. Rigorous analysis was provided in~\cite{MatveevTeimooriSavkin09,MatveevTeimooriSavkin11} to demonstrate the efficacy of the proposed control law when the UAV does not stay close to the target initially. Due to the nature of local stability, it is necessary to design a controller that guarantees global convergence. To overcome the local stability issue, a ``aiming'' controller using both range and range rate measurements was developed in~\cite{CaoMCK13}. The aiming controller is to control the heading of the UAV such that the UAV moves towards a well-designed circle. Global stability was shown in~\cite{CaoMCK13} regardless of the initial state of the UAV. Although range measurement is possible under a GPS-denied environment~\cite{SahinogluGenzici06}, the need of range rate measurement in the controller design in~\cite{CaoMCK13} is impractical as this measurement is typical unavailable or otherwise suffers from large errors. To further remove range rate measurement required in the controller design, the objective of this paper is to develop a new controller using range-only measurement such that circumnavigation is achieved under a GPS-denied environment by expanding on the preliminary work reported in~\cite{Cao14}.


The objective of the paper is fulfilled via a two-step analysis. First, a control algorithm based on range and range rate measurements is proposed to accomplish the circumnavigation mission. One promising feature of the control algorithm is that the associated control input is always bounded. Second, a sliding-mode estimator using range measurement is designed to accurately estimate range rate in finite time when applying the proposed control algorithm with range rate measurement being replaced by the estimated value. The design of this control algorithm is motivated by the study on sliding-mode observer in~\cite{DavilaFridmanLevant05,MorenoOsorio08,MorenoOsorio12}, where numerous velocity observers were designed based on position information. By combining the two steps, the circumnavigation mission can be accomplished using the proposed control algorithm when actual range rate measurement is replaced by its estimated value obtained from the sliding-mode estimator. To our best knowledge, this is the first paper that solves the circumnavigation problem using range-only measurement.

The rest of the paper is structured as follows. Section~\ref{sec:problem} introduces the circumnavigation problem. In Section~\ref{sec:new-controller}, a control algorithm based on range and range rate measurements is proposed to solve the circumnavigation problem, where the associated control input is always bounded. It is shown that UAV can always circumnavigate the target at the desired radius regardless of its initial state. In Section~\ref{sec:observed-range-rate}, range rate measurement used in the control algorithm in Section~\ref{sec:new-controller} is replaced by its estimated value obtained via a sliding-mode estimator. Such a replacement is valid because the estimated value and the actual value become identical after a finite period of time. In Section~\ref{sec:simulation}, two illustrative simulation examples are provided as a proof of concept. Finally, Section~\ref{sec:conclusion} is given to conclude the paper.

\section{Problem Statement}\label{sec:problem}
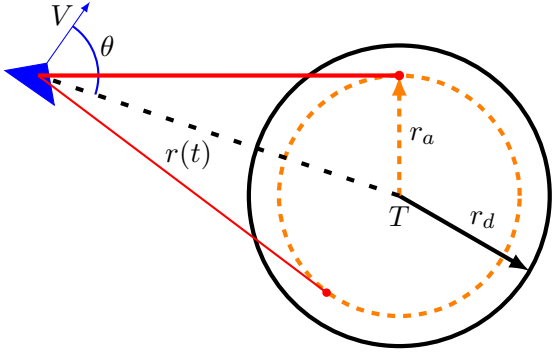
\begin{figure}
\centering
\begin{tikzpicture}[scale=.8]

\coordinate (center) at (8,3);
\coordinate (uav) at (2,5);
\draw [black, ultra thick] (center) circle [radius = 2.5];
\draw [orange, dashed, ultra thick] (center) circle [radius = 2.0];

\draw [-latex, black, ultra thick] (center) -- +(-30 : 2.5);
\draw[-latex, orange, dashed, ultra thick] (center) -- +(90:2);
\node [right] at (9,2.6) {$r_{d}$};
\node [right] at (8,4) {$r_{a}$};
\node [below] at (center) {$T$};

\draw [blue, fill=blue, rotate=-35, shift=(uav)] ++(90:.25) -- ++(-45:.707) -- ++(180:1) -- ++(45:.707);
\draw [black, loosely dashed, ultra thick] (center) -- (uav);
\node [below] at (4.5,4.1) {$r(t)$};
\draw [-latex, blue] (uav) -- ++(55:1.5);
\draw [blue,thick] (uav) +(55:1) arc [radius=1, start angle=55, end angle=-18.435];
\node [] at (3.15,5.5) {$\theta$};
\node [] at (2.4,6) {$V$};

\draw [red, ultra thick] (uav) -- (8,5);
\draw [red, fill=red] (8,5) circle [radius=2pt];
\draw [red, thick] (uav) -- +(-37:6) [red, fill=red] circle [radius=1.5pt];

\end{tikzpicture}
\caption{An illustrative example of some variables used in this paper. The blue triangle denotes the UAV. $T$ denotes the target. The blue arrow denotes the heading of the UAV. $r(t)$ denotes the range between the UAV and the target. $V$ denotes the (constant) velocity of the UAV. $\theta$ denotes the bearing angle. $r_d$ denotes the desired radius. $r_a$ is some design parameter in~\eqref{eq:controller-new}. }
\label{fig:motivation}
\end{figure}

Circumnavigation concerns with the behavior that a UAV orbits around an unknown target at some desired distance. For instance, in Fig.~\ref{fig:motivation}, let $T$ denote the unknown target whose location is $[x_T,y_T]^T$ and the blue triangle denote the UAV. The objective is to have the UAV orbit around $T$ at some desired distance $r_d$. In other words, the desired UAV trajectory is the black solid circle regardless of a counter clockwise or clockwise motion.

Assuming that UAV can maintain its altitude, we consider the following UAV dynamics given by
\begin{align}\label{eq:dynamics-UAV}
\dot{x}&=V\cos(\psi)\notag\\
\dot{y}&=V\sin(\psi)\\
\dot{\psi}&=\omega,\notag
\end{align}
where $[x,y]^T$ denotes the 2D location of the UAV, $\psi$ denotes the heading of the UAV, $\omega$ is the control input to be designed, and $V$ is the (constant) velocity of the UAV. Although this is a simplified model, it serves as a good approximation of practical UAV dynamics. Let the range measurement be denoted by $r\defeq \sqrt{(x-x_T)^2+(y-y_T)^2}$. The objective is to design control input $\omega$ such that $r(t)\to r_d$ as $t\to\infty$. Such a motion that UAV orbits around a target at a fixed radius is referred to as a stable circular motion, which is defined as follows.

\begin{definition}
A stable circular motion refers to the behavior that the UAV, with dynamics~\eqref{eq:dynamics-UAV},
moves around a target with a constant speed and a constant radius.
\end{definition}

Notice that~\eqref{eq:dynamics-UAV} characterizes the relationship between the controller $\omega$ and the state of the UAV $[x,y,\psi]$. Since the objective is to control $r$ by designing an appropriate controller $\omega$, it is thus desirable that~\eqref{eq:dynamics-UAV} can be rewritten into a new form, in which the relationship between $\omega$ and $r$ can be explicitly identified. Clearly, $r$ needs to be a state variable in the new form. Another variable used in the new form is the bearing angle, defined as follows.
\begin{definition}\label{de:bearing}
Denote the reference vector as the vector from the current location of the UAV to $T$. The bearing angle $\theta(t)\in[0,2\pi)$ at time $t$ is defined as the angle from the reference vector to the current heading of the UAV measured counterclockwise.
\end{definition}

As seen in Fig.~\ref{fig:motivation}, if the bearing angle is $\pi/2$ or $3\pi/2$, the heading of UAV is perpendicular to the vector from the UAV to the target. Physically, this means that the UAV cannot get closer to or farther away from the target. Indeed, if $r$ and $\theta$ are chosen as the state variables, the dynamics~\eqref{eq:dynamics-UAV} can be rewritten as
\begin{align}\label{eq:UAV-polar}
\dot{r}&=-V\cos(\theta)\notag\\
\dot{\theta}&=\omega+\frac{V\sin(\theta)}{r}.
\end{align}
The first equation in~\eqref{eq:UAV-polar}, illustrating the relationship between the bearing angle and the rate of $r$, matches the analyzed physical property.
To distinguish~\eqref{eq:dynamics-UAV} and~\eqref{eq:UAV-polar}, we call~\eqref{eq:dynamics-UAV} ``Cartesian dynamics'' and~\eqref{eq:UAV-polar} ``Polar dynamics''. Considering the objective of the paper, \textit{i.e.,} to design $\omega$ based on $r$ such that $r(t)\to r_d$ as $t\to \infty$,~\eqref{eq:UAV-polar} is used when referring to UAV dynamics although the two dynamics (Cartesian dynamics and Polar dynamics) are physically equivalent. As can be noticed from~\eqref{eq:UAV-polar}, the rate of $r$ is controlled by the bearing angle $\theta$, which can then be controlled by $\omega$. Intuitively, it is possible to design a controller $\omega$ to meet the desired property. However, it remains unclear if such a controller exists when only range measurement is available.

\section{A Controller Based on Range and Range Rate Measurements} \label{sec:new-controller}
In this section, a control algorithm based on range and range rate measurements is proposed to accomplish the circumnavigation mission. We show that the proposed control algorithm is able to guarantee globally asymptotic convergence and the equilibrium is exponentially stable. Globally asymptotic convergence means that the desired orbit motion is always guaranteed for an arbitrary initial state. System with exponentially convergence property has the benefit of improved robustness against disturbances.

Following the idea behind the control algorithm design in~\cite{CaoMCK13}, a new control algorithm is proposed as
\begin{equation}\label{eq:controller-new}
\omega=\left\{
\begin{array} {ll}
k[V\cos(\pi-\sin^{-1}(\frac{r_a}{r(t)}))-\dot{r}(t)],&r(t)\geq r_a,\\
0,&\text{otherwise},
\end{array}\right.
\end{equation}
where $k$ is a constant gain and $r_a$ is a parameter to be determined later. Notice that~\eqref{eq:controller-new} is a switching control law. To understand how the controller works, let's refer to Fig.~\ref{fig:motivation} for an illustrative situation. When $r(t)\geq r_a$, \textit{i.e.,} the UAV is on or outside the red dashed circle, the control input is determined by the difference between $V\cos(\pi-\sin^{-1}(\frac{r_a}{r(t)}))$ and $\dot{r}(t)$. One may consider $V\cos(\pi-\sin^{-1}(\frac{r_a}{r(t)}))$ as a reference for $\dot{r}(t)$ to track. In fact, $V\cos(\pi-\sin^{-1}(\frac{r_a}{r(t)}))$ refers to the change rate of $r(t)$ when the UAV moves towards one of the two tangent points on the red dashed circle. It should be emphasized that the two tangent points are not static as the UAV moves with a nonzero velocity. When $r(t)< r_a$, \textit{i.e.,} the UAV is inside the red dashed circle, the control input is zero, meaning that the UAV will move forward along its current heading. Because the UAV has a nonzero (constant) velocity, it takes a finite period of time before it exits the red dashed circle. Zero control when $r(t)< r_a$ is used to drive the UAV outside the undesirable zone while the feedback control $k[V\cos(\pi-\sin^{-1}(\frac{r_a}{r(t)}))-\dot{r}(t)]$ when $r(t)\geq r_a$ is to drive the UAV in such a way that the desired stable circular motion is reached eventually.

For sake of conciseness, we adopt the following definition even if it is a slight abuse of notation.
\begin{definition}\label{de:C_a}
The circle centered at $T$ with a radius $r_a$ is defined as $C_a$. The UAV is inside (resp., outside) $C_a$ if $r(t)<r_a$ (resp., $r(t)\geq r_a$).
\end{definition}

According to Definition~\ref{de:C_a}, the red dashed circle is $C_a$ in Fig.~\ref{fig:motivation}. As mentioned earlier, the radius of $C_a$, \textit{i.e.,} $r_a$, is to be designed. Before choosing a proper $r_a$, let's first analyze the radius of the stable circular motion under the proposed control algorithm~\eqref{eq:controller-new} assuming that such a stable circular motion does exist.

\begin{lemma}\label{lem:stable_radius}
Consider system dynamics~\eqref{eq:dynamics-UAV} subject to control input~\eqref{eq:controller-new}. If a stable circular motion exists, the radius is given by $r^\star=\sqrt{r_a^2+\frac{1}{k^2}}$. In addition, the UAV rotates clockwise (resp., counter clockwise) when $k>0$ (resp. $k<0$).
\end{lemma}
\proof When a stable circular motion exists, let the radius of the stable circular motion be given by $r^\star$. Then the magnitude of the nominal angular velocity is given by $\frac{V}{r^\star}$. Because the angular velocity of the UAV is equal to the control input~\eqref{eq:controller-new}, the magnitude of the nominal angular velocity is also equal to $\abs{\omega}$. For a stable circular motion, $\theta=\frac{\pi}{2}$ or $\theta=\frac{3\pi}{2}$, which implies that
\begin{itemize}
\item[(i)] The UAV cannot be inside $C_a$ because otherwise the UAV moves along a straight line;
\item[(ii)] $\dot{r}=0$ based on~\eqref{eq:UAV-polar}.
\end{itemize}
Then $\abs{\omega}$ becomes $\abs{kV\cos(\pi-\sin^{-1}(\frac{r_a}{r^\star}))}$ for the stable circular motion. It then follows that $\frac{V}{r^\star}$ and $\abs{kV\cos(\pi-\sin^{-1}(\frac{r_a}{r^\star}))}$ should be identical, which happens if and only if $r^\star=\sqrt{r_a^2+\frac{1}{k^2}}$.

When $k>0$ and $r(t)=r^\star$, it can be computed that $\omega<0$, indicating that the UAV rotates clockwise. When $k<0$, one can obtain that $\omega>0$, indicating that the UAV rotates counter clockwise.
\endproof

Lemma~\ref{lem:stable_radius} illustrates the relationship between the radius of the stable circular motion, if existing, and the parameter $r_a$ in~\eqref{eq:controller-new}. By choosing
\begin{equation}\label{eq:r_a}
r_a = \sqrt{r_d^2-\frac{1}{k^2}},
\end{equation}
one can obtain that $r^\star = r_d$ according to Lemma~\ref{lem:stable_radius}. One may observe that $r_a$ is strictly smaller than $r_d$, meaning that the UAV should aim towards a circle with a smaller radius in order to establish the desired circular motion. This is due to the nonlinear dynamics~\eqref{eq:UAV-polar}. For a nonnegative $r_a$, one can obtain that $k\geq \frac{1}{r_d}$.

The validity of Lemma~\ref{lem:stable_radius} is based on the assumption that a stable circular motion does exist. It remains an open question whether this assumption is true. Our effort next is to show that the assumption is indeed true. For the simplicity of presentation, we only consider the case when $k>0$. A similar analysis can be applied to the case when $k<0$.

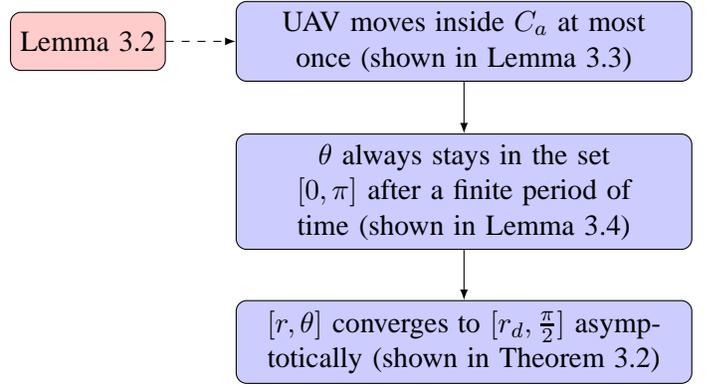
\begin{figure}
\centering
\tikzstyle{decision} = [rectangle, draw, fill=blue!20,
    text width=4.5em, text badly centered, node distance=5cm, inner sep=0pt]
\tikzstyle{block} = [rectangle, draw, fill=blue!20,
    text width=15em, text centered, rounded corners, minimum height=2em]
\tikzstyle{line} = [draw, -latex]
\tikzstyle{cloud} = [draw, rectangle, rounded corners, fill=red!20, node distance=5cm,
    minimum height=2em]
\begin{tikzpicture}[node distance = 2cm, auto]
    \node [block] (init) {UAV moves inside $C_a$ at most once (shown in Lemma~\ref{lem:2})};
    \node [cloud, left of=init] (expert) {Lemma~\ref{lem:1}};
    \node [block, below of=init] (evaluate) {$\theta$ always stays in the set $[0,\pi]$ after a finite period of time (shown in Lemma~\ref{lem:3})};
    \node [block, below of=evaluate] (stable) {$[r,\theta]$ converges to $[r_d,\frac{\pi}{2}]$ asymptotically (shown in Theorem~\ref{th:stability-new-controller})};
    \path [line] (init) -- (evaluate);
    \path [line] (evaluate) -- (stable);
    \path [line,dashed] (expert) -- (init);
\end{tikzpicture}
\caption{Three phases to prove that a stable circular motion exists using the control algorithm~\eqref{eq:controller-new}. }
\label{fig:3phase}
\end{figure}

Our focus next is to show that a stable circular motion is guaranteed using the control algorithm~\eqref{eq:controller-new} with $k>0$. As shown in Fig.~\ref{fig:3phase}, we arrange the proof based on three different phases. Phase 1 is to show that UAV moves inside $C_a$ at most once. The detailed analysis is shown in Lemma~\ref{lem:2} based on Lemma~\ref{lem:1}. Phase 2 is to show that $\theta$ always stays in the set $[0,\pi]$ after a finite period of time. The detailed analysis is shown in Lemma~\ref{lem:3}. Phase 3 is to show that $[r,\theta]$ converges to $[r_d,\frac{\pi}{2}]$ asymptotically. The detailed analysis is shown in Theorem~\ref{th:stability-new-controller}. Following this logic, we next present these three lemmas as well as the main theorem. The proofs of these lemmas and theorem are similar to those in~\cite{CaoMCK13} with some variations. To make the paper self-contained, complete proofs of these lemmas and theorem are provided next.

\begin{lemma}\label{lem:1}
Consider the UAV dynamics in~\eqref{eq:dynamics-UAV} subject to the control policy in~\eqref{eq:controller-new}. Let there be $t_0\geq 0$ such that $r(t_0) \geq r_a$ and $\theta(t_0)\in(\sin^{-1}(\frac{r_a}{r(t_0)}),2\pi-\sin^{-1}(\frac{r_a}{r(t_0)}))$, then $r(t) \geq r_a,$ $\forall t \geq t_0$.
\end{lemma}
\proof The proof of the lemma can be divided into the following two steps:

Step 1: $\theta(t)\in[\sin^{-1}(\frac{r_a}{r(t)}),2\pi-\sin^{-1}(\frac{r_a}{r(t)}))$ holds for all $t\geq t_0$. Based on the control algorithm~\eqref{eq:controller-new}, $\omega<0$ at $t=t_0$ since $\dot{r}(t)>V\cos(\pi-\sin^{-1}(\frac{r_a}{r(t)}))$ when $t=t_0$. As a consequence, the UAV will rotate clockwise initially. Note that both $\dot{r}(t)$ and $V\cos(\pi-\sin^{-1}(\frac{r_a}{r(t)}))$ are continuous with respect to $t$. Therefore, $\dot{r}(t)>V\cos(\pi-\sin^{-1}(\frac{r_a}{r(t)}))$ always holds before $\dot{r}(t)=V\cos(\pi-\sin^{-1}(\frac{r_a}{r(t)}))$ happens. Consequently, the UAV will stop rotating clockwise once $\dot{r}(t)=V\cos(\pi-\sin^{-1}(\frac{r_a}{r(t)}))$. Indeed, $\dot{r}(t)=V\cos(\pi-\sin^{-1}(\frac{r_a}{r(t)}))$ if and only if $\theta(t)=\sin^{-1}(\frac{r_a}{r(t)})$ or $\theta(t)=2\pi-\sin^{-1}(\frac{r_a}{r(t)})$. Thus, $\theta(t)\in[\sin^{-1}(\frac{r_a}{r(t)}),2\pi-\sin^{-1}(\frac{r_a}{r(t)})]$. To prove Step 1, it suffices to show that $\theta(t)=2\pi-\sin^{-1}(\frac{r_a}{r(t)})$ cannot hold. When $\theta(t_0)\in(\sin^{-1}(\frac{r_a}{r(t_0)}),2\pi-\sin^{-1}(\frac{r_a}{r(t_0)}))$ and $\omega= 0$, $\theta(t)=2\pi-\sin^{-1}(\frac{r_a}{r(t)})$ never holds for all $t\geq t_0$ because the UAV moves along a straight line and the straight line is always outside the circle $C_a$. As a consequence, $\theta(t)=2\pi-\sin^{-1}(\frac{r_a}{r(t)})$ never holds for all $t\geq t_0$ when $\theta(t_0)\in[\sin^{-1}(\frac{r_a}{r(t_0)}),2\pi-\sin^{-1}(\frac{r_a}{r(t_0)}))$ and $\omega\leq 0$. Combining the previous arguments completes the proof of Step 1.

Step 2: $r(t)\geq r_a$ for all $t\geq t_0$. When $r(t)=r_a$, it follows that $\sin^{-1}(\frac{r_a}{r(t)})=\frac{\pi}{2}$. From Step 1, it is known that $\theta(t)\in[\sin^{-1}(\frac{r_a}{r(t)}),2\pi-\sin^{-1}(\frac{r_a}{r(t)}))$. This means that $\theta(t)\in(\frac{\pi}{2},\frac{3\pi}{2}]$ at the time when $r(t)=r_a$. By recalling the second equation in~\eqref{eq:UAV-polar},
one can obtain that $\dot{r}\geq 0$ at the time when $r(t)=r_a$. This implies that the UAV cannot get any closer to the target if $r(t)=r_a$. At the time $t^\star$ when $r(t^\star)$ becomes larger than $r_a$, the bearing angle has to be in the set $(\frac{\pi}{2},\frac{3\pi}{2})$, which satisfies the condition that $\theta(t^\star)\in[\sin^{-1}(\frac{r_a}{r(t^\star)}),2\pi-\sin^{-1}(\frac{r_a}{r(t^\star)}))$. By repeating the analysis in Steps 1 and 2, it is clear that the UAV will never move inside $C_a$.
\endproof


\begin{lemma}\label{lem:2}
Consider the UAV dynamics in~\eqref{eq:dynamics-UAV} subject to the control policy in~\eqref{eq:controller-new}. The UAV can only move inside $C_a$ at most once.
\end{lemma}
\proof When the UAV enters $C_a$ at some time $t_e$, \textit{i.e.,} $r(t_e)=r_a$, $\theta(t_e)\in[0,\frac{\pi}{2})\cup(\frac{3\pi}{2},2\pi)$ holds true in order to guarantee that the UAV enters $C_a$. Recall that no control input is imposed on the UAV when it is inside $C_a$. Then at the time $t_x$ when it exits $C_a$,
\begin{align*}
\theta(t_x)=
\left\{\begin{array} {ll}
\pi-\theta(t_e),&\theta(t_e)\in[0,\frac{\pi}{2}),\\
3\pi-\theta(t_e),&\theta(t_e)\in(\frac{3\pi}{2},2\pi),
\end{array}\right.
\end{align*}
because the UAV moves along a straight line due to the fact $\omega=0$. As a consequence, $\theta(t_x)\in(\frac{\pi}{2},\frac{3\pi}{2})$. When $r(t_x)=r_a$, it follows that $\sin^{-1}(\frac{r_a}{r(t_x)})=\frac{\pi}{2}$. Therefore, $\theta(t_x)\in(\sin^{-1}(\frac{r_a}{r(t_x)}),2\pi-\sin^{-1}(\frac{r_a}{r(t_x)}))$ when $r(t_x)=r_a$. By considering the current time $t_x$ be the time $t_0$ in Lemma~\ref{lem:1}, it follows from Lemma~\ref{lem:1} that the UAV will never move inside $C_a$ again. Therefore, the UAV can only move inside $C_a$ at most once.
\endproof

\begin{lemma}\label{lem:3}
Consider the UAV dynamics in~\eqref{eq:dynamics-UAV} subject to the control policy in~\eqref{eq:controller-new}. For any $\theta(0)$, there exists $t^\star\geq 0$ such that $\theta(t)\in[0,\pi]$ for any $t\geq t^\star$.
\end{lemma}
\proof By Lemma~\ref{lem:2}, the UAV can move inside $C_a$ at most once. When the UAV never moves inside $C_a$, let $t_1=0$. When the UAV moves inside $C_a$ once, let $t_1$ be the time when the UAV moves from inside $C_a$ to outside $C_a$. It is clear that $t_1$ is finite. The lemma is proved if the following two statements are valid:\newline
(1) For any $\theta(0)$, there exists $t^\star\geq t_1$ such that $\theta(t^\star)\in[0,\pi]$; and\newline
(2) Once $\theta(t^\star)\in[0,\pi]$ for some $t^\star\geq t_1$, $\theta(t)\in[0,\pi]$ for any $t\geq t^\star$.

The first statement is proved by considering the following three cases:
\begin{itemize}
\item[(i)] $\theta(t_1)\in(\pi,2\pi-\sin^{-1}(\frac{r_a}{r(t_1)}))$: From~\eqref{eq:UAV-polar}, one can obtain that $\dot{\theta}(t)<0$, \textit{i.e.,} $\theta(t)$ will decrease, whenever $\theta(t)\in(\pi,2\pi-\sin^{-1}(\frac{r_a}{r(t)}))$ because $\omega<0$ while $-\frac{V\sin(\theta(t))}{r(t)}>0$. When $\omega$ is (approximately) zero, $-\frac{V\sin(\theta(t))}{r(t)}$ is (approximately) $\frac{Vr_a}{r^2(t)}$. Because both $\omega$ and $-\frac{V\sin(\theta(t))}{r(t)}$ are continuous with respect to $t$, $\dot{\theta}(t)$ is always upper bounded by some negative constant. Similarly, when $-\frac{V\sin(\theta(t))}{r(t)}$ is (approximately) zero, $\omega$ is upper bounded by some negative constant, indicating that $\dot{\theta}(t)$ is always upper bounded by some negative constant as well. Therefore, $\theta(t)\leq \pi$ in finite time. Noting that $\theta(t_1)\in(\pi,2\pi-\sin^{-1}(\frac{r_a}{r(t_1)}))\in(\sin^{-1}(\frac{r_a}{r(t_1)}),2\pi-\sin^{-1}(\frac{r_a}{r(t_1)}))$, it follows from Step 1 in the proof of Lemma~\ref{lem:1} that $\theta(t)$ cannot get smaller than $\sin^{-1}(\frac{r_a}{r(t)})$, which implies that $\theta(t)\geq \sin^{-1}(\frac{r_a}{r(t)})$.
\item[(ii)] $\theta(t_1)=2\pi-\sin^{-1}(\frac{r_a}{r(t_1)})$: Under this case, the UAV is heading towards the tangent point such that the bearing is $2\pi-\sin^{-1}(\frac{r_a}{r(t_1)})$. By computation, $\omega=0$, which implies that the UAV will move along a straight line towards the tangent point. As a consequence, the UAV will move towards the tangent point whenever $r(t)>r_a$. Given a constant nonzero velocity, it takes a finite period of time before $r(t)=r_a$ happens. Notice that $r(t)=r_a$ cannot hold for an arbitrary period of time because otherwise a contradiction happens by noting that (i) $\omega=0$ based on~\eqref{eq:controller-new} during that period of time, indicating that the UAV cannot rotate; and (ii) the UAV has to rotate such that $r(t)=r_a$ holds for that period of time. For $t\geq t_1$, the UAV cannot move inside $C_a$, as discussed in the first paragraph of the proof. This implies that $r(t)$ will increase to be greater than $r_a$ as soon as $r(t)=r_a$ happens. The bearing angle $\theta(t_f)$ must be in the interval $(\frac{\pi}{2},\frac{3\pi}{2})$ at the time when $r(t)$ increases to be greater than $r_a$. When $\theta(t_f) \in(\pi,\frac{3\pi}{2})$, it follows from Case (i) that $\theta(t)$ will be in the set $[0,\pi]$ after a finite period of time. When $\theta(t_f) \in(\frac{\pi}{2},\pi]$, it is already in the set $[0,\pi]$.
\item[(iii)] $\theta(t_1)=(2\pi-\sin^{-1}(\frac{r_a}{r(t_1)}),2\pi)$: If $\theta(t)=(2\pi-\sin^{-1}(\frac{r_a}{r(t)}),2\pi)$ always holds, it takes a finite period of time before $r(t)\leq r_a$ happens. Since the UAV never moves inside $C_a$ for $t\geq t_1$, either Case (i) or Case (ii) will happen after a finite period of time. By following the analysis in Cases (i) and (ii), $\theta(t)$ will be in the set $[0,\pi]$ after a finite period of time.
\end{itemize}
To prove the second statement, it is essential to study $\dot{\theta}(t)$ when $\theta(t)=0$ or $\theta(t)=\pi$.
Because the UAV is outside $C_a$, it can be computed that
\[\omega=k[V\cos(\pi-\sin^{-1}(\frac{r_a}{r(t)}))+V]>0,\]
where the first equation in~\eqref{eq:UAV-polar} was used to derive the equality. This indicates that $\theta(t)$ will increase as soon as $\theta(t)=0$ happens. Similarly, when $\theta(t)=\pi$, one can obtain that
\[\omega=k[V\cos(\pi-\sin^{-1}(\frac{r_a}{r(t)}))-V]<0,\]
which indicates that $\theta(t)$ will decrease as soon as $\theta(t)=\pi$ happens.  Therefore the second statement holds as well.
\endproof

It can be noted that the proof of Lemma~\ref{lem:2} depends on Lemma~\ref{lem:1} and the proof of Lemma~\ref{lem:3} depends on Lemma~\ref{lem:2}. With these three lemmas, we now present the main result in this section.

\begin{theorem}\label{th:stability-new-controller}
Consider the UAV dynamics in~\eqref{eq:dynamics-UAV} subject to the control policy in~\eqref{eq:controller-new}. If $k>\frac{1}{r_d}$ and $r_a$ is chosen satisfying~\eqref{eq:r_a}, then $r(t) \to r_d$ and $\theta(t)\to\frac{\pi}{2}$ as $t \to \infty$.
\end{theorem}
\proof According to Lemmas~\ref{lem:2} and~\ref{lem:3}, there exists a time instant $t^\star$ such that $r(t)\geq r_a$ and $\theta(t)\in[0,\pi]$. Then the control input can be simplified as
\begin{equation*}
\omega=k[V\cos(\pi-\sin^{-1}(\frac{r_a}{r(t)}))+V\cos(\theta)],\quad t\geq t^\star.
\end{equation*}
For $t\geq t^\star$, consider a Lyapunov function candidate given by
\begin{align*}
\Vcal = 1-\sin(\theta)+\varphi,
\end{align*}
where
$\varphi=\int_{r_d}^r (\frac{1}{r_d}-\frac{1}{z}+k\cos\sin^{-1}(\frac{r_a}{z})-k\cos\sin^{-1}(\frac{r_a}{r_d}))\text{d}z\geq 0.$ Because
\[
\frac{1}{r_d}-\frac{1}{z}\left\{
\begin{array} {ll}
>0,&z>r_d,\\
=0,&z=r_d,\\
<0,&z<r_d,
\end{array}\right.
\]
and $k\cos\sin^{-1}(\frac{r_a}{z})-k\cos\sin^{-1}(\frac{r_a}{r_d})$ satisfies such a property as well, combining with the property of integration shows that $\varphi\geq 0$, indicating that $\Vcal\geq 0$ as $1-\sin(\theta)\geq 0$. When $r_a$ is chosen satisfying~\eqref{eq:r_a}, one can obtain that $\frac{1}{r_d}=k\cos\sin^{-1}(\frac{r_a}{r_d})$ by computation. Then $\varphi$ can be simplified as $\int_{r_d}^r (-\frac{1}{z}+k\cos\sin^{-1}(\frac{r_a}{z}))\text{d}z$. Taking derivative of $\Vcal$ yields that
\begin{align*}
\dot{\Vcal}&=-\cos(\theta)\dot{\theta}+\dot{\varphi}\\
&=-\cos(\theta)\Bigg[kV\left(-\cos\sin^{-1}(\frac{r_a}{r})+\cos(\theta)\right)+\frac{V\sin(\theta)}{r}\Bigg]\\
&~~~-\left[k\cos\sin^{-1}(\frac{r_a}{r})-\frac{1}{r}\right]V\cos(\theta)\\
&=V\cos(\theta)\left[-k\cos(\theta)-\frac{\sin(\theta)}{r}+\frac{1}{r}\right],
\end{align*}
where~\eqref{eq:UAV-polar} was used to derive the second equality. When $\theta\in[\frac{\pi}{2},\pi]$, $\dot{\Vcal}\leq 0$ because $\cos(\theta)\leq 0$ and $-k\cos(\theta)-\frac{\sin(\theta)}{r}+\frac{1}{r}\geq 0$. When $\theta\in[0,\frac{\pi}{2})$, $\dot{\Vcal}\leq 0$ because $\cos(\theta)> 0$ and $-k\cos(\theta)-\frac{\sin(\theta)}{r}+\frac{1}{r}\leq -\frac{\cos(\theta)}{r}-\frac{\sin(\theta)}{r}+\frac{1}{r}\leq 0$. Therefore, $\dot{\Vcal}\leq 0$. Note that $\dot{\Vcal}$ is uniformly continuous when $r\geq r_a$ and $\theta\in[0,\pi]$, it follows from Lemma 4.3 in~\cite{SlotineLi91} that $\dot{\Vcal}\to 0$ as $t\to\infty$. When $k>\frac{1}{r_d}$, $\dot{\Vcal}=0$ implies that $\theta=\frac{\pi}{2}$. It then follows from~\eqref{eq:UAV-polar} that when $\theta(t)=\frac{\pi}{2}$, $r(t)$ is constant. Therefore, a stable circular motion does exist. It then follows from the analysis in the paragraph right after Lemma~\ref{lem:1} that $r(t)=r_d$ if $r_a$ is chosen satisfying~\eqref{eq:r_a}. Therefore, $\theta(t)\to\frac{\pi}{2}$ and $r(t)\to r_d$ as $t\to\infty$.
\endproof

\begin{myremark}
Although it is assumed that the velocity of the UAV, $V$, is constant, Lemmas~\ref{lem:stable_radius},~\ref{lem:1},~\ref{lem:2},~\ref{lem:3}, and Theorem~\ref{th:stability-new-controller} are still valid if $V$ is varying within a set $[\underline{V},\overline{V}]$, where $\underline{V}$ and $\overline{V}$ are two positive constant. In other words, the assumed constant velocity for the UAV is not required as long as the velocity is both lower bounded and upper bounded. Lemma~\ref{lem:stable_radius} also shows that the final stable radius remains unchanged even if the velocity of the UAV changes.
\end{myremark}

In the previous part of this section, the proposed controller~\eqref{eq:controller-new} was shown to guarantee global asymptotic stability regardless of the initial state. By observation, one can find that the control input associated with~\eqref{eq:controller-new} is always bounded by $2kV$ because $\abs{\cos(\pi-\sin^{-1}(\frac{r_d}{r(t)}))}$ is bounded by $1$ and $\abs{\dot{r}}\leq V$ due to~\eqref{eq:UAV-polar}. Although this controller~\eqref{eq:controller-new} uses both range and range rate measurements, we will show in the next section that the requirement of range rate measurement can be eliminated thanks to the boundedness of the control input.

Before proceeding to the next section, we show that the closed-loop system of~\eqref{eq:UAV-polar} subject to~\eqref{eq:controller-new} is exponentially stable at its equilibrium. If the equilibrium is exponentially stable, the associated closed-loop system not only converges, but in fact converges at a rate faster or at least as fast as some know rate near the equilibrium. One benefit of such a system is its robustness against disturbances around the equilibrium.

\begin{theorem}
Consider the closed-loop system given by~\eqref{eq:UAV-polar} subject to~\eqref{eq:controller-new}. If $k>\frac{1}{r_d}$ and $r_a$ is chosen satisfying~\eqref{eq:r_a}, then $[r_d,\frac{\pi}{2}]^T$ is a locally exponentially stable equilibrium.
\end{theorem}
\proof For the UAV dynamics in~\eqref{eq:UAV-polar} subject to the control policy in~\eqref{eq:controller-new}, the corresponding closed-loop system can be written as
\begin{align}
\dot{r}(t) =& -V\cos(\theta(t))\label{eq:rdot-rewrite}\\
\dot{\theta}(t)=&k[V\cos(\pi-\sin^{-1}(\frac{r_a}{r(t)}))+V\cos(\theta(t))]\notag\\
&+\frac{V\sin(\theta(t))}{r(t)}, \label{eq:theta-dot-rewrite}
\end{align}
where~\eqref{eq:rdot-rewrite} was substituted into the first equation in~\eqref{eq:UAV-polar} to obtain~\eqref{eq:theta-dot-rewrite}.
Let's define
\[ f_1(r(t),\theta(t))\defeq-V\cos(\theta(t))\]
and
\begin{align*}
f_2(r(t),\theta(t))\defeq &k[V\cos(\pi-\sin^{-1}(\frac{r_a}{r(t)}))+V\cos(\theta(t))]\\
&+\frac{V\sin(\theta(t))}{r(t)}.
\end{align*}
The linearization of~\eqref{eq:rdot-rewrite} and~\eqref{eq:theta-dot-rewrite} around the equilibrium $[r_d,\frac{\pi}{2}]^T$ is given by
\begin{equation}
\dot{x}(t) = A(t)x(t),
\end{equation}
where $x(t)=[r(t),\theta(t)]^T$ and $A(t)\defeq [A_{ij}]\in\re^{2\times 2}$ with
\begin{align*}
A_{11} = \frac{\partial f_1(r(t),\theta(t))}{\partial r(t)}\mid_{r(t)=r_d,\theta(t)=\frac{\pi}{2}}=0
\end{align*}
\begin{align*}
A_{12}= \frac{\partial f_1(r(t),\theta(t))}{\partial \theta(t)}\mid_{r(t)=r_d,\theta(t)=\frac{\pi}{2}}=V
\end{align*}
\begin{align*}
A_{21}=& \frac{\partial f_2(r(t),\theta(t))}{\partial r(t)}\mid_{r(t)=r_d,\theta(t)=\frac{\pi}{2}}\\=&-\frac{kVr_a^2}{r_d^3\sqrt{1-\frac{r_a^2}{r_d^2}}}-\frac{V}{r_d^2}
\end{align*}
and
\begin{align*}
A_{22}= \frac{\partial f_2(r(t),\theta(t))}{\partial \theta(t)}\mid_{r(t)=r_d,\theta(t)=\frac{\pi}{2}}=-kV.
\end{align*}
Then the eigenvalues of $A(t)$ are given by $$\frac{-kV\pm \sqrt{(kV)^2-V(\frac{kVr_a^2}{r_d^3\sqrt{1-\frac{r_a^2}{r_d^2}}}+\frac{V}{r_d^2})}}{2}.$$
Clearly, $A(t)$ is Hurwitz. It then follows from~\cite{Khalil02} that the equilibrium $[r_d,\frac{\pi}{2}]$ is exponentially stable.
\endproof


\section{A Revised Controller Based on Range Measurement and Estimated Range Rate} \label{sec:observed-range-rate}
In Section~\ref{sec:new-controller}, a controller based on range and range rate measurements was presented to guarantee the desired circular motion for the UAV. Direct range rate measurement is typically unavailable or suffers from significant uncertainties. The purpose of this section is to remove range rate measurement needed in the control algorithm~\eqref{eq:controller-new}. In particular, an estimated range rate, obtained via a sliding-mode estimator using range measurement, is used to replace the range rate measurement used in~\eqref{eq:controller-new}.

Here control algorithm~\eqref{eq:controller-new} is revised as
\begin{equation}\label{eq:controller-obs}
\widehat{\omega}=\left\{
\begin{array} {ll}
k[V\cos(\pi-\sin^{-1}(\frac{r_a}{r(t)}))-\hat{x}_2],&r(t)\geq r_a,\\
0,&\text{otherwise},
\end{array}\right.
\end{equation}
where $\hat{x}_2$ is an estimate of $\dot{r}$. In particular, $\hat{x}_2$ is obtained via the following sliding-mode estimator as
\begin{align}\label{eq:smo}
\dot{\hat{x}}_1&=\hat{x}_2+k_1\abs{r-\hat{x}_1}^{\frac{1}{2}}\text{sgn}(r-\hat{x}_1),\notag\\
\dot{\hat{x}}_2&=k_2\text{sgn}(r-\hat{x}_1)+k_3(r-\hat{x}_1),
\end{align}
if $r\geq r_a$ and
\begin{align}\label{eq:smo-zero}
\dot{\hat{x}}_1&=0,\quad \hat{x}_1(t_x)=2r_d-\hat{x}_1(t_e)\notag\\
\dot{\hat{x}}_2&=0, \quad \hat{x}_2(t_x)=-\hat{x}_2(t_e),
\end{align}
if $r<r_a$, where $\text{sgn}(\cdot)$ is the sign function, $t_e$ is the time when UAV moves inside $C_a$ from outside $C_a$\footnote{If the UAV is initially inside $C_a$, the initial time is one $t_e$.} and $t_x$ denotes the first subsequent time when the UAV moves outside $C_a$, and $k_i, i=1,2,3,$ are positive constants. Because the UAV could move inside and outside $C_a$ multiple times, multiple $t_e$ and $t_x$ may be expected. In fact, the number of $t_e$ and the number of $t_x$ should be exactly the same since the UAV cannot stabilize inside $C_a$ due to the zero control applied when the UAV is inside $C_a$. Here $\hat{x}_1$ can be considered an estimate of $r$. Briefly speaking, the main idea behind the estimator is that (i) $\hat{x}_1$ and $\hat{x}_2$ satisfy~\eqref{eq:smo} if the UAV is outside $C_a$; (ii) $\hat{x}_1$ and $\hat{x}_2$ remain unchanged if the UAV is inside $C_a$; and (iii) once the UAV moves outside $C_a$, $\hat{x}_2$ is reset as its negate while $\hat{x}_1$ is reset as $2r_d$ plus its negate. As shown in the proof of the following Theorem~\ref{th:observer-circum}, the reset of $\hat{x}_1$ and $\hat{x}_2$ at the time when the UAV moves from inside $C_a$ to outside $C_a$ is crucial in establishing a finite-time convergence of $(\hat{x}_1,\hat{x}_2)$ to $(r,\dot{r})$.

\begin{myremark}
Instead of using actual range rate in controller~\eqref{eq:controller-new}, estimated range rate is used in controller~\eqref{eq:controller-obs}.
Because the designed sliding mode estimator guarantees that
the estimation error converges to zero in finite time, the wellknown
“separation principle” can be applied in the controller
design. That is, the design of range rate estimator and the
design of a feedback controller based on estimated range rate
can be decoupled into two separated problems.
\end{myremark}

Before presenting the main result in this section, the following lemma is needed.

\begin{lemma}\label{lem:estimator}
Consider the differential equation given by
\begin{align}
\dot{p}&=q-k_1\abs{p}^{\frac{1}{2}}\text{sgn}(p),\notag\\
\dot{q}&=-k_2\text{sgn}(p)-k_3p+f(t,p,q),\label{eq:lemma-diff-equation}
\end{align}
where $\abs{f(t,p,q)}<\delta_1+\delta_2 \abs{q}$ with $\delta_i>0,~i=1,2$. If $k_1>0$, $k_2>\max\{1+\frac{\delta_1^2}{k_1},\frac{1}{2}\delta_2^2+2\delta_2\}$, and $k_3>0$, $(p,q)$ approaches $(0,0)$ in finite time.
\end{lemma}
\proof Let $\xi\defeq [\abs{p}^{\frac{1}{2}}\text{sgn}(p),p,q]^T$ and consider the following Lyapunov function candidate given by
\begin{equation}\label{eq:Lyapunov}
\Vcal = \xi^TP\xi,
\end{equation}
where \[P=\frac{1}{2}\left[\begin{array}{ccc}
4k_2+k_1^2 & 0 & -k_1\\
0 & 2k_3 & 0\\
-k_1 & 0 & 2\end{array}\right].\] Note that $\Vcal(x)$ is positive-definite and is differentiable almost everywhere except at $p=0$. When $p\neq 0$, the derivative of $\Vcal$ is given by $\dot{\Vcal}=\xi^TP\dot{\xi}$. Notice that the derivative of $\xi$ is given by $[\frac{1}{2}\abs{p}^{-\frac{1}{2}}\dot{p},\dot{p},\dot{q}]^T$. By recalling~\eqref{eq:lemma-diff-equation}, $\dot{\Vcal}$ can be rewritten as
\begin{align}\label{eq:Vdot}
\dot{\Vcal}=&-\abs{p}^{-\frac{1}{2}}\xi^TQ_1\xi-\xi^TQ_2\xi-k_1f(t,p,q)\abs{p}^{\frac{1}{2}}\text{sgn}(p)\notag\\&+2qf(t,p,q),
\end{align}
where \[Q_1=\frac{k_1}{2}\left[\begin{array}{ccc}
2k_2+k_1^2 & 0 & -k_1\\
0 & 2k_3 & 0\\
-k_1 & 0 & 1\end{array}\right]\] and \[Q_2=k_2\left[\begin{array}{ccc}
k_2+2k_1^2 & 0 & 0\\
0 & k_4 & 0\\
0 & 0 & 1\end{array}\right].\]

Because $\abs{f(t,p,q)}\leq \delta_1+\delta_2 q$ under the assumption of the lemma, it can be further obtained that
\begin{align*}
&\abs{k_1f(t,p,q)\abs{p}^{\frac{1}{2}}\text{sgn}(p)}\\
\leq &\delta_1k_1\abs{p}^{\frac{1}{2}}+k_1\delta_2\abs{q}\abs{p}^{\frac{1}{2}}\\
=&\delta_1k_1\abs{p}^{-\frac{1}{2}}[\abs{p}^{\frac{1}{2}}\text{sgn}(p)]^2+\frac{1}{2}\{\delta_2^2q^2+k_1^2[\abs{p}^{\frac{1}{2}}\text{sgn}(p)]^2\}
\end{align*}
and
\begin{align*}
&\abs{2qf(t,p,q)}\\
\leq &2\delta_1 \abs{q}+2\delta_2 q^2\\
=&2\delta_1 \abs{p}^{-\frac{1}{2}}\abs{\abs{p}^{\frac{1}{2}}\text{sgn}(p)}\abs{q}+2\delta_2 q^2\\
\leq&\delta_1 \abs{p}^{-\frac{1}{2}}\{[\abs{p}^{\frac{1}{2}}\text{sgn}(p)]^2+q^2\}+2\delta_2 q^2.
\end{align*}
Then it follows from~\eqref{eq:Vdot} that
\begin{align*}
\dot{\Vcal}\leq &-\abs{p}^{-\frac{1}{2}}\xi^TQ_1\xi-\xi^TQ_2\xi+\abs{p}^{-\frac{1}{2}}\xi^TQ_3\xi+\xi^TQ_4\xi,
\end{align*}
where
\[Q_3=\left[\begin{array}{ccc}
\delta_1(1+k_1) & 0 & 0\\
0 & 0 & 0\\
0 & 0 & \delta_1\end{array}\right]\]
and
\[Q_4=\left[\begin{array}{ccc}
\frac{1}{2}k_1^2 & 0 & 0\\
0 & 0 & 0\\
0 & 0 & \frac{1}{2}\delta_2^2+2\delta_2\end{array}\right].\]
When $k_i,~i=1,2,3,$ satisfy the conditions in the lemma, $Q_1-Q_3$ and $Q_2-Q_4$ are positive definite. It then can be obtained that
\begin{align*}
\dot{\Vcal}\leq &-\abs{p}^{-\frac{1}{2}}\xi^T(Q_1-Q_3)\xi\\
\leq&-\norm{\xi}^{-\frac{1}{2}}\xi^T(Q_1-Q_3)\xi\\
\leq&-\norm{\xi}^{-\frac{1}{2}}\lambda_{\min}(Q_1-Q_3)\norm{\xi}^2\\
=&-\lambda_{\min}(Q_1-Q_3)\norm{\xi},
\end{align*}
where $\lambda_{\min}(\cdot)$ denotes the minimum eigenvalue of a symmetric positive-definite matrix, $\abs{p}\leq \norm{\xi}$ is used to derive the second inequality, $\xi^T(Q_1-Q_3)\xi\geq \lambda_{\min}(Q_1-Q_3)\norm{\xi}^2$ is used to derive the third inequality. Because $\Vcal\leq \lambda_{\max}(P)\norm{\xi}^2$, where $\lambda_{\max}(\cdot)$ denotes the maximum eigenvalue of a symmetric positive-definite matrix, it follows that $\norm{\xi}\geq \frac{1}{\sqrt{\lambda_{\max}(P)}}\Vcal^{\frac{1}{2}}$. It can then be obtained that
\begin{align*}
\dot{\Vcal}\leq -\frac{\lambda_{\min}(Q_1-Q_3)}{\sqrt{\lambda_{\max}(P)}}\Vcal^{\frac{1}{2}}.
\end{align*}
After some manipulation, one can get that
\begin{align}\label{eq:V-ineq}
2\sqrt{\Vcal(t)}\leq 2\sqrt{\Vcal(0)}-\frac{\lambda_{\min}(Q_1-Q_3)}{\sqrt{\lambda_{\max}(P)}}t.
\end{align}

Because $\Vcal$ is differentiable almost everywhere except at $p=0$,~\eqref{eq:V-ineq} is valid almost everywhere except at $p=0$. Note that $\Vcal=0$ if and only if $p=0$ and $q=0$. Therefore, $\Vcal(t)\to 0$ in finite time, which implies that $(p,q)$ approaches $(0,0)$ in finite time.
\endproof

\begin{myremark}
In~\cite{MorenoOsorio08}, finite-time convergence of~\eqref{eq:lemma-diff-equation} with $f(t,p,q)\leq \delta_1+\delta_2\abs{p}$ was studied. Lemma~\ref{lem:estimator} further analyzed the case when $f(t,p,q)\leq \delta_1+\delta_2\abs{q}$.
\end{myremark}

With Lemma~\ref{lem:estimator}, we are now ready to present the main result in this section.

\begin{theorem}\label{th:observer-circum}
Consider the UAV dynamics in~\eqref{eq:dynamics-UAV} subject to the control policy in~\eqref{eq:controller-obs}. If $k>\frac{1}{r_a}$, $k_1>0$, $k_2>\max\{1+\frac{V^4(2k+\frac{1}{r_a})^2}{k_1},\frac{1}{2}k^2V^2+2kV\}$ and $k_3>0$, then $r(t) \to r_d$ and $\theta(t)\to\frac{\pi}{2}$ as $t \to \infty$, where $r_a=\sqrt{r_d^2-\frac{1}{k^2}}$.
\end{theorem}
\proof Notice from Theorem~\ref{th:stability-new-controller} that $r(t) \to r_d$ and $\theta(t)\to\frac{\pi}{2}$ as $t \to \infty$ by using~\eqref{eq:controller-new}, where $r_a=\sqrt{r_d^2-\frac{1}{k^2}}$. Then it is sufficient to prove this theorem if~\eqref{eq:controller-obs} and~\eqref{eq:controller-new} become identical after a finite period of time because one can simply consider the initial time be the time after which~\eqref{eq:controller-obs} and~\eqref{eq:controller-new} are always identical. Therefore, our focus next is to show~\eqref{eq:controller-obs} and~\eqref{eq:controller-new} become identical in finite time under the condition of the theorem.

Define $x_1\defeq r$ and $x_2\defeq \dot{r}$. By recalling the polar dynamics in~\eqref{eq:UAV-polar} and the control policy~\eqref{eq:controller-obs}, the derivatives of $x_1$ and $x_2$ are given by
\begin{align}\label{eq:example}
\dot{x}_1&=x_2,\notag\\
\dot{x}_2&=V\sin(\theta)\left[\widehat{\omega}+\frac{V\sin(\theta)}{x_1}\right].
\end{align}
Let $p\defeq x_1-\hat{x}_1$ and $q\defeq x_2-\hat{x}_2$. When $r\geq r_a$, \textit{i.e.,} $x_1\geq r_a$, it follows from~\eqref{eq:example} and~\eqref{eq:smo} that
\begin{align*}
\dot{p}=&q-k_1\abs{p}^{\frac{1}{2}}\text{sgn}(p),\\
\dot{q}=&V\sin(\theta)\left[\widehat{\omega}+\frac{V\sin(\theta)}{x_1}\right]-k_2\text{sgn}(p)-k_3p\\
=&\underbrace{V\sin(\theta)\left[-kV\cos\sin^{-1}(\frac{r_a}{r})-kx_2-kq+\frac{V\sin(\theta)}{x_1}\right]}_{f(t,p,q)}\\
&-k_2\text{sgn}(p)-k_3p.
\end{align*}
Because $x_2=\dot{r}=-V\cos(\theta)$, one can obtain that
\begin{align*}
\abs{f(t,p,q)}\leq V(2kV+k\abs{q}+\frac{V}{r_a}).
\end{align*}
By letting $\delta_1=V^2(2k+\frac{1}{r_a})$ and $\delta_2=kV$, $\abs{f(t,p,q)}\leq\delta_1+\delta_2\abs{q}$. The property of $f(t,p,q)$ in Lemma~\ref{lem:estimator} is fulfilled under the condition of the theorem. By considering the Lyapunov function $\Vcal$ defined in~\eqref{eq:Lyapunov}, it follows from the analysis in Lemma~\ref{lem:estimator} that
\begin{align}\label{eq:der-property}
\dot{\Vcal}\leq -\eta \Vcal^{\frac{1}{2}}
\end{align}
holds almost everywhere for some positive $\eta$ that is determined by $k$, $k_1$, $k_2$, and $k_3$ under the condition of the theorem.

Now let's consider the case when $r< r_a$, \textit{i.e.,} $x_1<r_a$. Because $\widehat{\omega}=0$, function $f(t,p,q)$ becomes $\frac{[V\sin(\theta)]^2}{x_1}$, which is not necessarily bounded since $x_1$ might approach zero. The property $\abs{f(t,p,q)}\leq\delta_1+\delta_2\abs{q}$ for some positive $\delta_1$ and $\delta_2$ is not necessarily satisfied. To overcome this issue, let's drop the time when the UAV is inside $C_a$. Because zero control input is imposed when the UAV is inside $C_a$ and $V>0$, it takes a finite period of time before the UAV moves outside $C_a$. Without loss of generality, let $t_e$ be the time when the UAV moves inside $C_a$ and $t_x$ be the first subsequent time when the UAV moves outside $C_a$. Clearly, $t_x-t_e$ is bounded (by $\frac{2r_a}{V}$). By only considering the case when the UAV is outside $C_a$, the time interval $[t_e,t_x)$ is dropped. One consequence is that the state $(p,q)$ might be different at time instants $t_e$ and $t_x$. The change of $(p,q)$ at the two time instants could result in a jump of $\Vcal$ from $t_e$ to $t_x$. If such a jump exists, it is desired that $\Vcal$ becomes smaller or at least remains unchanged. Then the Lyapunov function still satisfies~\eqref{eq:der-property} almost everywhere and jumps to be smaller at some distinct instants if the time interval $[t_e,t_x)$ is dropped.

When the UAV is initially outside $C_a$, let the initial time be unchanged. When the UAV is initially inside $C_d$, let the initial time be the time when the UAV moves outside $C_d$ for the first time.
Notice that $\Vcal$ in~\eqref{eq:Lyapunov} can be written as
\begin{align}\label{eq:V-rewritten}
\Vcal= &2k_2\abs{p}+k_3p^2+\frac{1}{2}q^2+\frac{1}{2}\left[k_1\abs{p}^{\frac{1}{2}}\text{sgn}(p)-q\right]^2.
\end{align}
When $p(t_x)=2r_d-\hat{x}_1(t_e)$ and $q(t_x)=-\hat{x}_2(t_e)$, it can be computed that
\begin{align*}
p(t_x)=&x_1(t_x)-\hat{x}_1(t_x)\\
=&x_1(t_x)-[2r_d-\hat{x}_1(t_e)]\\
=&r_d-[2r_d-\hat{x}_1(t_e)]\\
=&\hat{x}_1(t_e)-r_d\\
=&-p(t_e).
\end{align*}
Recall from~\eqref{eq:dynamics-UAV} that $\dot{r}=-V\cos(\theta)$ and from the proof of Lemma~\ref{lem:2} that $\theta(t_x)=\pi-\theta(t_e)$ or $\theta(t_x)=3\pi-\theta(t_e)$. One can obtain that $\dot{r}(t_x)=-\dot{r}(t_e)$. Equivalently, $x_2(t_x)=-x_2(t_e)$. It thus follows that
\begin{align*}
q(t_x)=&x_2(t_x)-\hat{x}_2(t_x)\\
=&-x_2(t_e)-[-\hat{x}_2(t_x)]\\
=&-q(t_e).
\end{align*}
From~\eqref{eq:V-rewritten}, it can be observed that $\Vcal$ remains unchanged if both $p$ and $q$ are negated. Thus, $\Vcal$ remains unchanged when the state $(p(t_e),q(t_e))$ changes to $(p(t_x),q(t_x))$ under the proposed control algorithm~\eqref{eq:controller-obs}. If the finite time interval $[t_e,t_x)$ is dropped, $\Vcal$ is continuous and satisfies~\eqref{eq:der-property} almost everywhere. By following the analysis in Lemma~\ref{lem:estimator}, $\Vcal$ approaches zero in finite time if $[t_e,t_x)$ is excluded. This implies that $\hat{x}_1-r$ and $\hat{x}_2-\dot{r}$ go to zero in finite time if $[t_e,t_x)$ is excluded. In other words, there exists a time instant $t^\star$ such that $r(t^\star)\geq r_a$, $\hat{x}_1(t)-r(t)=0$, and $\hat{x}_2(t)-\dot{r}(t)=0$ for any $t\geq t^\star$ if $[t_e,t_x)$ is excluded. That is,~\eqref{eq:controller-obs} and~\eqref{eq:controller-new} are identical for $t\geq t^\star$ when $[t_e,t_x)$ is excluded. For $t\in[t_e,t_x)$,~\eqref{eq:controller-obs} and~\eqref{eq:controller-new} are identical because both of them are zero. Therefore,~\eqref{eq:controller-obs} and~\eqref{eq:controller-new} are always identical for all $t\geq t^\star$ if we consider $t^\star$ as the initial time.

By recalling the analysis in the first paragraph of the proof of the theorem, it can be obtained that $r(t) \to r_d$ and $\theta(t)\to\frac{\pi}{2}$ as $t \to \infty$.
\endproof

\begin{myremark}
One unique feature of the proposed control algorithm~\eqref{eq:controller-obs} is that only range measurement is needed in its implementation. Considering the limited sensing capabilities for small UAVs due to their payload restrictions, the control algorithm~\eqref{eq:controller-obs} and the concept behind it are more applicable in small UAV operations under a GPS-denied environment.
\end{myremark}


%

\section{Simulation Examples}\label{sec:simulation}
In this section, two simulation examples are provided to demonstrate the effectiveness of the two control algorithms proposed in Sections~\ref{sec:new-controller} and~\ref{sec:observed-range-rate}. For both control algorithms~\eqref{eq:controller-new} and~\eqref{eq:controller-obs}, the parameters are chosen as: $r_d=10$, $[x_T,y_T]=[0,-10]$, $k=0.2$, $V=1$. The initial state of the UAV is given by $[13,-2,5\pi/4]$. By computation, $r_a=\sqrt{r_d^2-\frac{1}{k^2}}=8.6603$. For control algorithm~\eqref{eq:controller-obs}, we further let $k_1 = 2$, $k_2 = 1.2$, $k_3 = 0.1$ and the sliding-mode estimator be initialized as $[10,0]$. These parameters are chosen in such a way that those conditions in Theorems~\ref{th:stability-new-controller} and~\ref{th:observer-circum} are satisfied.

\begin{figure}
\begin{center}
\includegraphics[width=.5\textwidth]{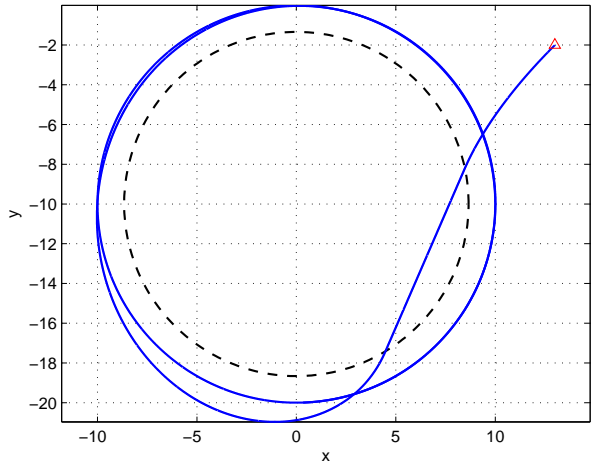}    
\caption{The trajectory of the UAV under~\eqref{eq:controller-new}. The red triangle represents the starting position of the UAV. The dashed circle represents $C_a$.}  
\label{fig:traj1}                                 
\end{center}                                 
\end{figure}

\begin{figure}
\begin{center}
\includegraphics[width=.5\textwidth]{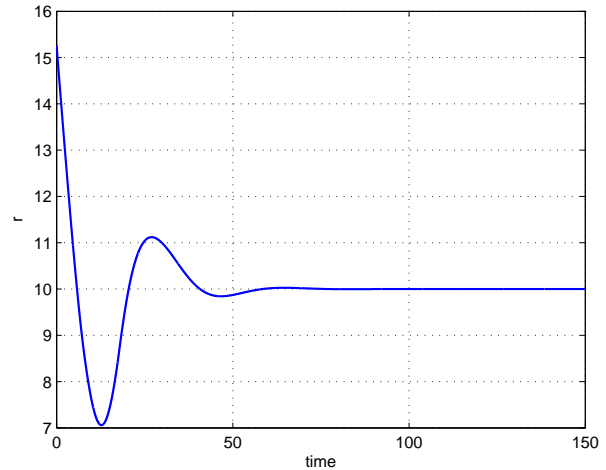}    
\caption{The distance between the UAV and the target under~\eqref{eq:controller-new}, \textit{i.e.,} $r(t)$. }  
\label{fig:range1}                                 
\end{center}                                 
\end{figure}

Figs.~\ref{fig:traj1} and~\ref{fig:range1} show, respectively, the trajectory of the UAV and the distance between the UAV and the target using control algorithm~\eqref{eq:controller-new}. Both figures show such a fact that the desired distance can be reached eventually. The fact that $r_a=8.6603$ means that the UAV aims towards a circle centered at the target's location with radius $8.6603$ in order to stabilize at the desired radius $10$.

\begin{figure}
\begin{center}
\includegraphics[width=.5\textwidth]{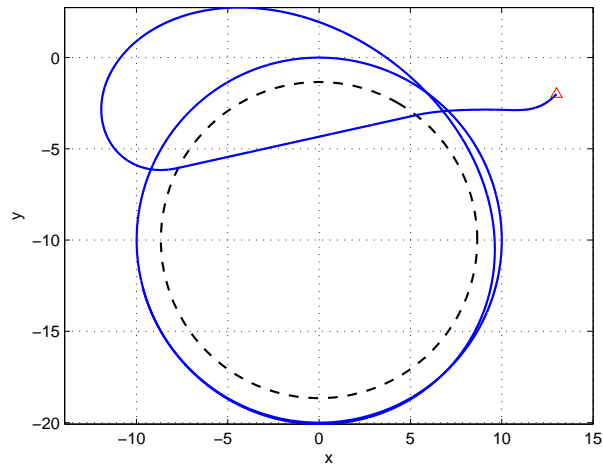}    
\caption{The trajectory of the UAV under~\eqref{eq:controller-obs}. The red triangle represents the starting position of the UAV. The dashed circle represents $C_a$.}  
\label{fig:traj}                                 
\end{center}                                 
\end{figure}

\begin{figure}
\begin{center}
\includegraphics[width=.5\textwidth]{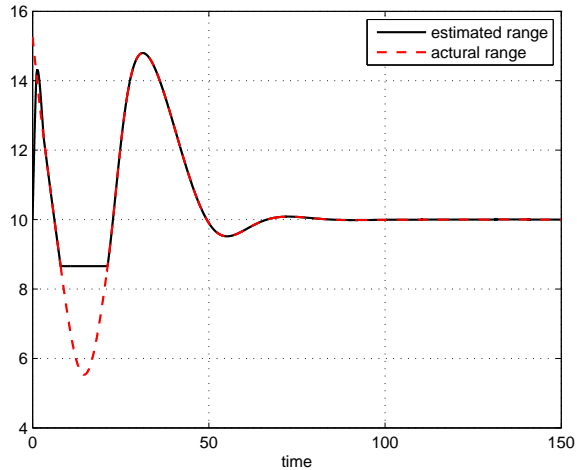}    
\caption{Estimated range $\hat{x}_1$ vs actual range $r$.}  
\label{fig:range}                                 
\end{center}                                 
\end{figure}

\begin{figure}
\begin{center}
\includegraphics[width=.5\textwidth]{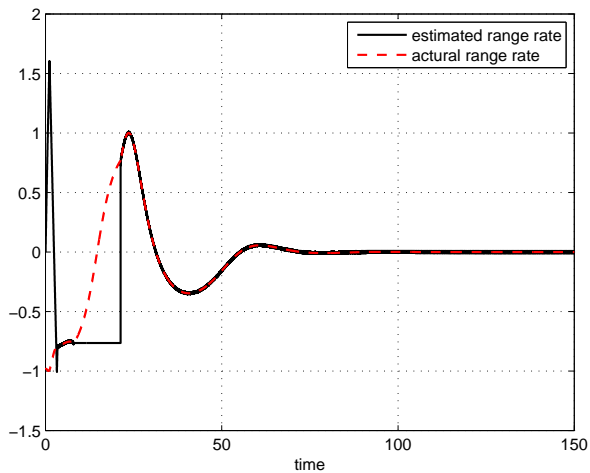}    
\caption{Estimated range rate $\hat{x}_2$ vs actual range rate $\dot{r}$.}  
\label{fig:rangerate}                                 
\end{center}                                 
\end{figure}

Figs.~\ref{fig:traj},~\ref{fig:range}, and~\ref{fig:rangerate} are the simulation results using control algorithm~\eqref{eq:controller-obs}. Fig.~\ref{fig:traj} shows the trajectory of the UAV under the control algorithm~\eqref{eq:controller-obs}, where the blue line represents the actual trajectory. Again, the desired distance between the UAV and the target can be reached eventually. Fig.~\ref{fig:range} shows the estimated range $\hat{x}_1$ and the actual range $r$ under the control algorithm~\eqref{eq:controller-obs}. It can be observed that the estimated range approaches the actual range in finite time. Fig.~\ref{fig:rangerate} shows the estimated range rate $\hat{x}_2$ and the actual range rate $\dot{r}$ under the control algorithm~\eqref{eq:controller-obs}. Again, the estimated range rate approaches the actual range rate in finite time if the time interval during which the UAV is inside $C_d$ is excluded. One can also see from Fig.~\ref{fig:rangerate} that the reset mechanism in~\eqref{eq:smo-zero} is necessary to guarantee the accurate estimate of $\dot{r}$ because otherwise the estimated range rate will be significantly different from the actual one at the time when the UAV exits $C_d$.

\section{Conclusion and Future Work}\label{sec:conclusion}
In this paper, we proposed a control algorithm based on range-only measurement such that a UAV can circumnavigate an unknown target at the desired distance under a GPS-denied environment. The design of such a control algorithm can be divided into two steps. First, a control algorithm based on range and range rate measurements was proposed to solve the circumnavigation problem. Second, a sliding-mode estimator was designed to replace the range rate measurement used in the control algorithm proposed in the first step. By choosing several parameters carefully, the sliding-mode estimator is able to accurately estimate the range rate in finite time using range measurement. Thus, the circumnavigation mission can be accomplished if the control algorithm in the first step is applied with range rate measurement being replaced by its estimated value obtained in the second step. Because the proposed control algorithm based on range-only measurement requires a minimum number of measurement for implementation, it is very suitable for UAV operations under a GPS-denied environment when limited measurements are available.

One future research direction is to study how measurement noises and wind affect the proposed control algorithm and test the control algorithm in real flights. Another interesting topic is to consider control input saturation since the allowable control input is normally limited for physical UAVs.

\section*{Acknowledgement}
The author would like to thank Prof. Brian Anderson for his insightful suggestions and comments.

\bibliographystyle{ieeetran}
\bibliography{refs}

\end{document}